\documentclass[12pt]{iopart}
\usepackage{iopams}  
\usepackage{graphicx}

\usepackage{bm}

\begin{document}
\title[Gain control in molecular signaling]{Gain control in molecular information processing: Lessons from
  neuroscience}

\author{Ilya Nemenman}
\address{Departments of Physics and Biology, Computational and Life Sciences Initiative\\ Emory University, Atlanta, GA 30322, USA}
\ead{ilya.nemenman@emory.edu}

\begin{abstract}
  Statistical properties of environments experienced by biological
  signaling systems in the real world change, which necessitates
  adaptive responses to achieve high fidelity information
  transmission. One form of such adaptive response is gain
  control. Here we argue that a certain simple mechanism of gain
  control, understood well in the context of systems neuroscience,
  also works for molecular signaling.  The mechanism allows to
  transmit more than one bit (on or off) of information about the
  signal independently of the signal variance.  It does not require
  additional molecular circuitry beyond that already present in many
  molecular systems, and, in particular, it does not depend on
  existence of feedback loops. The mechanism provides a potential
  explanation for abundance of ultrasensitive response curves in
  biological regulatory networks.
\end{abstract}

\date{\today}
\maketitle

\section{Introduction}
An important function of all biological systems is responding to
signals from the surrounding environment.  These signals (hereafter
assumed to be scalars), $s(t)$, are often probabilistic, described by
some probability distribution $P[s(t)]$.  They have non-trivial
temporal dynamics, so that the probability of a certain value of the
signal at a given time is dependent on its entire history.

Often the response $r(t)$ is produced from $s$ by (possibly nonlinear
and noisy) temporal filtering. For example, in a deterministic
molecular circuit, we may have
\begin{equation}
  \frac{dr}{dt}=f\left(s\left(t\right)\right)-kr,
\label{filter}
\end{equation}
where $f$ is the response molecule synthesis rate, which depends on
the current value of the signal. Here $k$ is the rate of the
first-order degradation of the molecule. Note that $r(t)$ depends on
the entire history of $s(t')$, $t'<t$, and hence carries information
about it. For more complicated, nonlinear degradation or for
$r$-dependent synthesis, Eq.~(\ref{filter}) may be interpreted as
linearization around the mean response.

\begin{figure}[b]
\centerline{\includegraphics[width = 8cm]{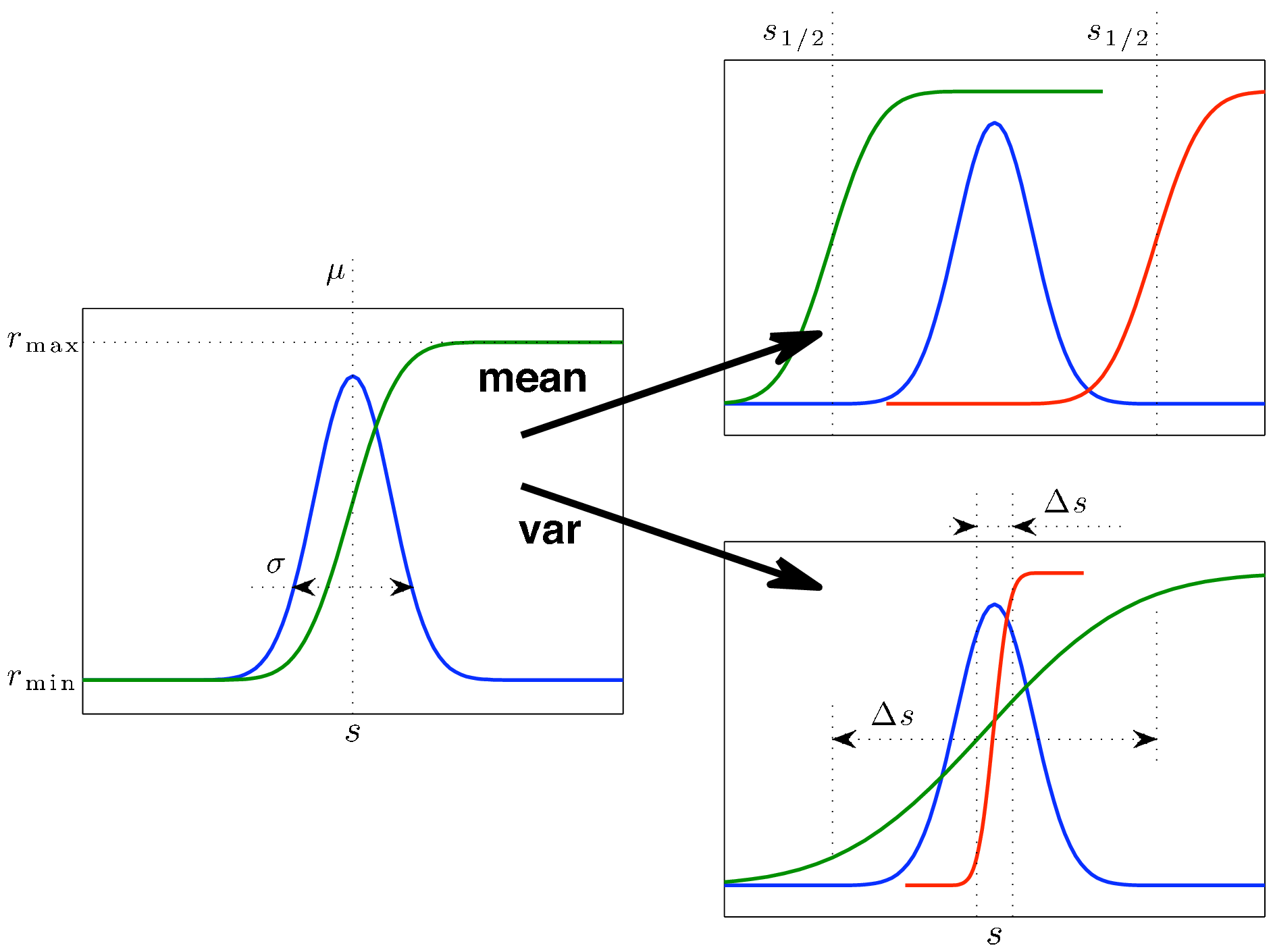}}
\caption{Parameters characterizing response to a signal. Left panel:
  the probability distribution of the signal, $P(s)$ (blue), and the
  best-matched steady state dose-response curve $r_{\rm ss}$
  (green). Top right: if the mid-point of the dose-response curve,
  $s_{1/2}$, is far away from the mean of the signal, a typical
  response will be extremal. Bottom right: if the width of the
  dose-response curve, $\Delta s$, is considerably different from the
  standard deviation of the signal, then the typical response is
  either extremal, or at its mid-point. These mismatches prevent using
  the entire dynamic range of the response to convey information about
  the signal.}
\label{response}
\end{figure}
The distribution of stimuli, $P[s(t)]$, places severe constraints on
admissible forms of $f$. To see this, for quasi-stationary signals
(that is, when the signal correlation time $\tau$ is large,
$\tau\gg1/k$), we use Eq.~(\ref{filter}) to write the steady state
dose-response curve
\begin{equation}
  r_{\rm ss}=f\left(s(t)\right)/k.
\label{dose-response}
\end{equation}
A typical monotonic, sigmoidal $f$ is characterized by only a few
large-scale parameters: the range, $[f_{\min}, f_{\max}]$; the
mid-point $s_{1/2}$; and the width of the transition region, $\Delta
s$ (cf.\ Fig.~\ref{response}). If the signal mean $\mu\gg s_{1/2}$,
then, for most signals, $r_{\rm ss}\approx f_{\rm max}/k$. Then
responses to two different signals $s_1$ and $s_2$ are
indistinguishable as long as
\begin{equation}
\left.\frac{dr_{\rm ss}(s)}{ds}\right|_{s=s_1}(s_2-s_1)<\delta r,
\end{equation}
where $\delta r$ is the precision of the response
resolution. Similarly, when $\mu\ll s_{1/2}$, then $r_{\rm ss}\approx
f_{\rm min}/k$. Thus, for reliably communicating information about the
signal, $f$ should be tuned such that $s_{1/2} \approx \mu$. If a
biological system can change its $s_{1/2}$ to follow changes in $\mu$,
this is called {\em adapting to the mean} of the signal, and, if
$s_{1/2}(\mu)=\mu$, then the adaptation is {\em perfect}
\cite{berg,nemenman-11}.  Similarly, if the quasi-stationary signal is
taken from the distribution with $\sigma\equiv \left(\langle
  s(t)^2\rangle _t-\mu^2\right)^{1/2}\gg \Delta s$, then the response
to most of the signals will be indistinguishable from the extrema. It
will be near $\sim (r_{\max}+r_{\min})/2$ if $\sigma\ll \Delta
s$. Thus, to use the full dynamic range of the response, a biological
system must tune the width of the sigmoidal dose-response curve to
$\Delta s\approx\sigma$; this is called the {\em variance adaptation}
or {\em gain control} \cite{nemenman-11}.

Both of these adaptation behaviors can be traced to the same
theoretical argument \cite{laughlin-81}: for sufficiently general
conditions on the response resolution $\delta r$, the response that
optimizes the fidelity of a signaling system, as measured by its
information-theoretic channel capacity \cite{shannon-49}, is $r_{\rm
  ss}^*(s)=\int_{-\infty}^sP(s')ds'$, where $P(s')$ is the probability
distribution of an instantaneous signal value, obtained from
$P[s(t)]$. However, since environmental changes that lead to varying
$\mu$ and $\sigma$, as well as mechanisms of the adaptation may be
distinct, it often makes sense to consider the two adaptations as
separate phenomena \cite{nemenman-11}.

Adaptation to the mean, sometimes also called {\em desensitization},
has been observed and studied in a wide variety of biological sensory
systems
\cite{berg,norman-perlman,laughlin-81,desense-molecular,detwiler-etal-00},
with active work persisting to date. In contrast, while gain control
has been investigated in neurobiology
\cite{brenner-00,adapt,borst-05}, we are not aware of its systematic
analysis in molecular sensing. In this article, we start filling in
the gap.  Our main contribution is the observation that a mechanism
for gain control, observed in a fly motion estimation system by Borst
et al.~\cite{borst-05}, can be transferred to molecular information
processing with minimal modifications. Importantly, unlike adaptation
to the mean, which is implemented typically using extra feedback
circuitry \cite{berg,detwiler-etal-00,chao}, the gain control
mechanism we analyse requires no additional regulation.  It is
built-in into many molecular signaling systems.  The main ingredients
of the gain control mechanism in Ref.~\cite{borst-05} is a strongly
nonlinear, sigmoidal response function $f(s)$ and a realization that
real-world signals are dynamic with a nontrivial temporal
structure. Thus one must move away from the steady state response
analysis and autocorrelations within the signals will allow the
response to carry more information about the signal than seems
possible naively.

Specifically, we show that even a simple biochemical circuit in
Eq.~(\ref{filter}), with no extra regulatory features can be made
insensitive to changes in $\sigma$. That is, for an arbitrary choice
of $\sigma$, and for a wide range of other parameters, the circuit can
generate an output that is informative of the input, and, in
particular, carries more than a single bit of information about it.
For brevity, we will not review the original work on gain control in
neural systems \cite{borst-05}, but will instead develop the
methodology directly in the molecular context.

\section{Results: Gain control with no additional regulatory
  structures}
Let's assume for simplicity that the signal in Eq.~(\ref{filter}) has
the Ornstein-Uhlenbeck dynamics with:
\begin{equation}
  \langle s(t)\rangle =\mu,\;\;\langle s(t+t')s(t)\rangle
  =\sigma^2e^{-t'/\tau}. \label{signal} 
\end{equation} 
We will assume that the response has been adapted to the mean value of
this signal (likely by additional feedback control circuitry, not
considered here explicitly), so that the response to $s=\mu$ is half
maximal. Now we explore how insensitivity to $\sigma$ can be achieved
as well.

\begin{figure}[b]
\centerline{\includegraphics[width = 8cm]{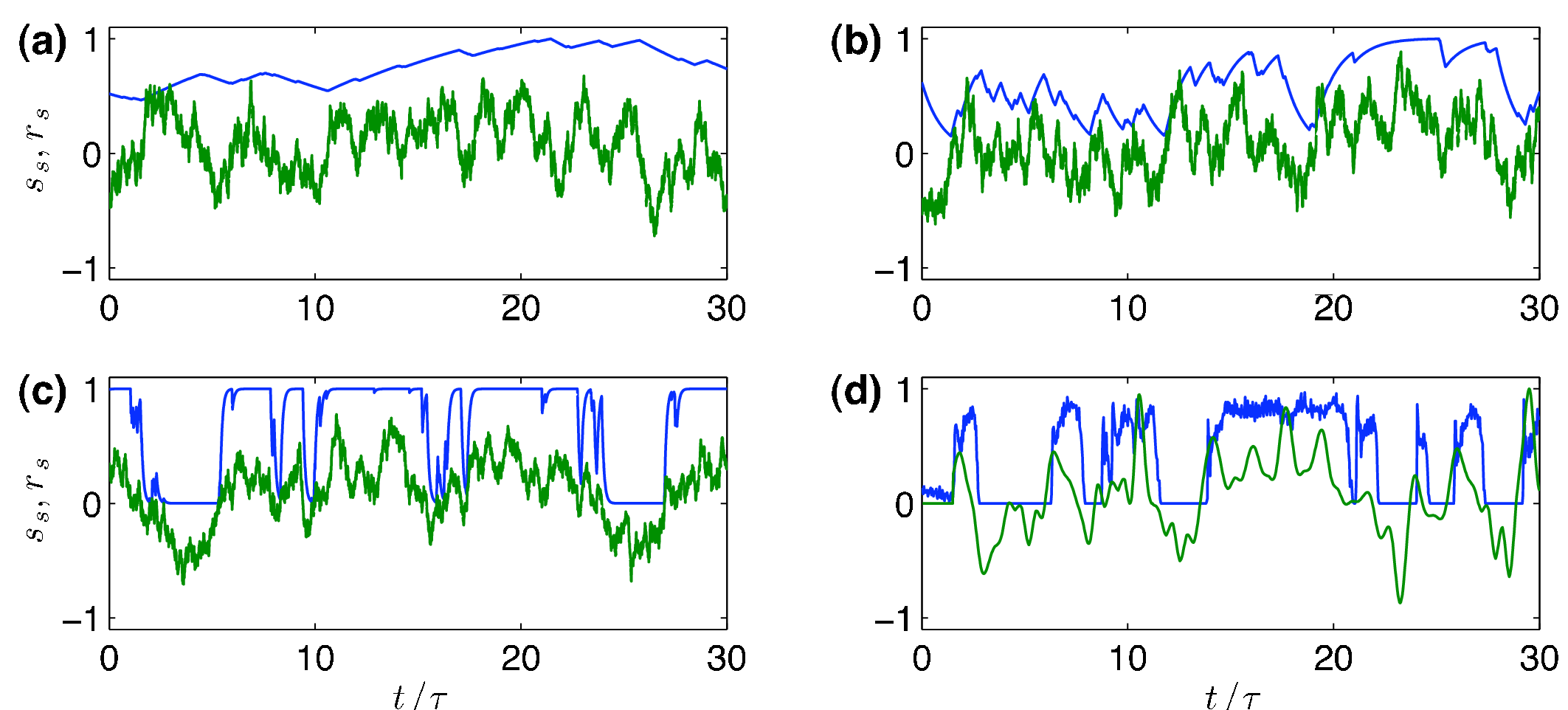}}
\caption{Examples of signals and responses for dynamics in
  Eqs.~(\ref{filter}, \ref{signal}, \ref{thresholdf}). On the vertical
  axis, we plot normalized signals $s_s=(s-m)/\max(s)$ (green) and
  normalized responses $r=r/\max(r)$ (blue). On the horizontal axis,
  the time is rescaled by the correlation time of the signal,
  $\tau$. Panels (a)-(c) have $k\tau=0.1,1,10$, respectively [recall
  that $1/k$ is the response time of the circuit,
  Eq.~(\ref{filter})]. In panel (d), we show, for comparison, the
  firing rate of a blow fly motion sensitive neuron H1 and its driving
  stimulus, both rescaled to one (see \cite{nemenman-etal-08} for
  details of this experiment). The stimulus had little power at high
  frequencies, but the single-exponential correlation structure held
  for long times. Notice the similarity between the
  telegraph-series-like structure of the responses in panels (c) and
  (d). Since the H1 neuron served as a model neural system for the
  feedback-free gain adaptation in \cite{borst-05}, this similarity
  suggests to look for gain-controlled responses in molecular
  signaling, Eq.~(\ref{filter}), as well. }
\label{examples}
\end{figure}

We start with a step-function approximation to the sigmoidal response
synthesis
\begin{equation}
  f=f_0\theta(s-\mu)=f_0\times\left\{
    \begin{array}{l}
      0,\; s<\mu,\\
      1/2,\; s=\mu',\\ 
      1,\; s>\mu,\end{array}\right.
\label{thresholdf}
\end{equation}
where $f_0$ is some constant. This is a limiting case of very high
Hill number dose-response curves, which have been observed in nature
\cite{cluzel-motor}.  Figure~\ref{examples} shows sample signals and
responses produced by this system. Notice that such $f$ makes the
system manifestly insensitive to $\sigma$. Any changes in $\sigma$
will not result in changes to the response, hence the gain is
controlled {\em perfectly}.

Nevertheless, this choice of $f$ is pathological, resulting in a
binary steady state response ($r_{\rm ss}=0$ for $s<\mu$, and $r_{\rm
  ss}=f_0/k$ otherwise). That is, the response cannot carry more than
one bit of information about the stimulus. However, as illustrated in
Fig.~\ref{examples}, a {\em dynamic} response is not binary and varies
over its entire dynamic range. Can this make a difference and produce
a dose-response relation that is both high fidelity and insensitive to
the variance of the signal?

\begin{figure}[b]
\centerline{\includegraphics[width = 6cm]{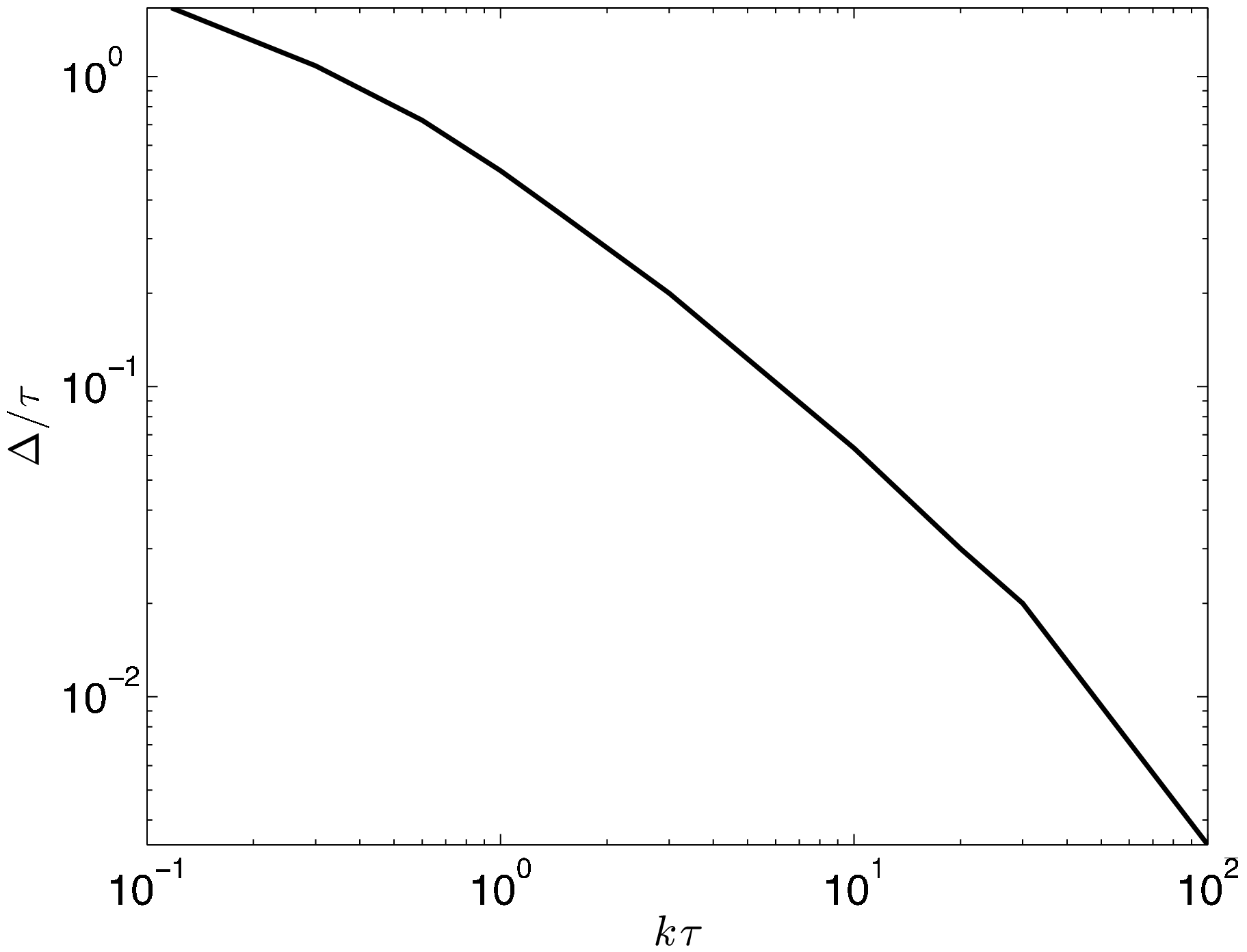}}
\caption{Dependence of the delay between the signal and the response,
  $\Delta$, which achieves the maximum correlation between $s$ and
  $r$. Here $\Delta$ is expressed in units of the signal correlation
  time $\tau$, and it is studied as a function of $k\tau$, the ratio
  of characteristic time scales of the signal and the response
  dynamics.}
\label{corr_delay}
\end{figure}

To answer this, we first specify what we mean by the dose-response
curve or the input-output relation when there is no steady state
response. For the response at a single time point $t$, we can write
$P\left(r\left(t\right)|\{s\left(t'\le
    t\right)\}\right)=\delta(r-r[s])$, where $\delta(\cdots)$ is the
Dirac $\delta$-function, and the functional $r[s]$ is obtained by
solving Eq.~(\ref{filter}).  Since the signal is probabilistic,
marginalizing over all but the instantaneous value of it at time
$t-t'$, one gets $P\left(r\left(t\right)|s\left(t-t'\right)\right)$,
the distribution of the response at time $t$ conditional on the value
of the signal at $t-t'$.  Further, for the distribution of the signal
given by Eq.~(\ref{signal}), one can numerically integrate
Eq.~(\ref{filter}) and evaluate the correlation $c(t')=\langle
r(t)s(t-t')\rangle_t$ \footnote{All simulations were performed using
  Matlab v.\ 7.6 and Octave v.\ 3.0.2 using Apple Macbook
  Air. Correlation time of the signal was $\tau=300$ integration time
  steps, and averages were taken over $3\times 10^6$ time steps. To
  change the value of $k\tau$, only $k$ was adjusted.  }. Since
Eq.~(\ref{filter}) is causal, $c(t')$ has a maximum at some
$t'=\Delta(\tau,k)\ge 0$, illustrated in
Fig.~\ref{corr_delay}. Correspondingly, in this paper we replace the
familiar notion of the dose-response curve by the probabilistic
input-output relation $P(r(t)|s(t-\Delta))$.

\begin{figure}[b]
\centerline{\includegraphics[width = 6cm]{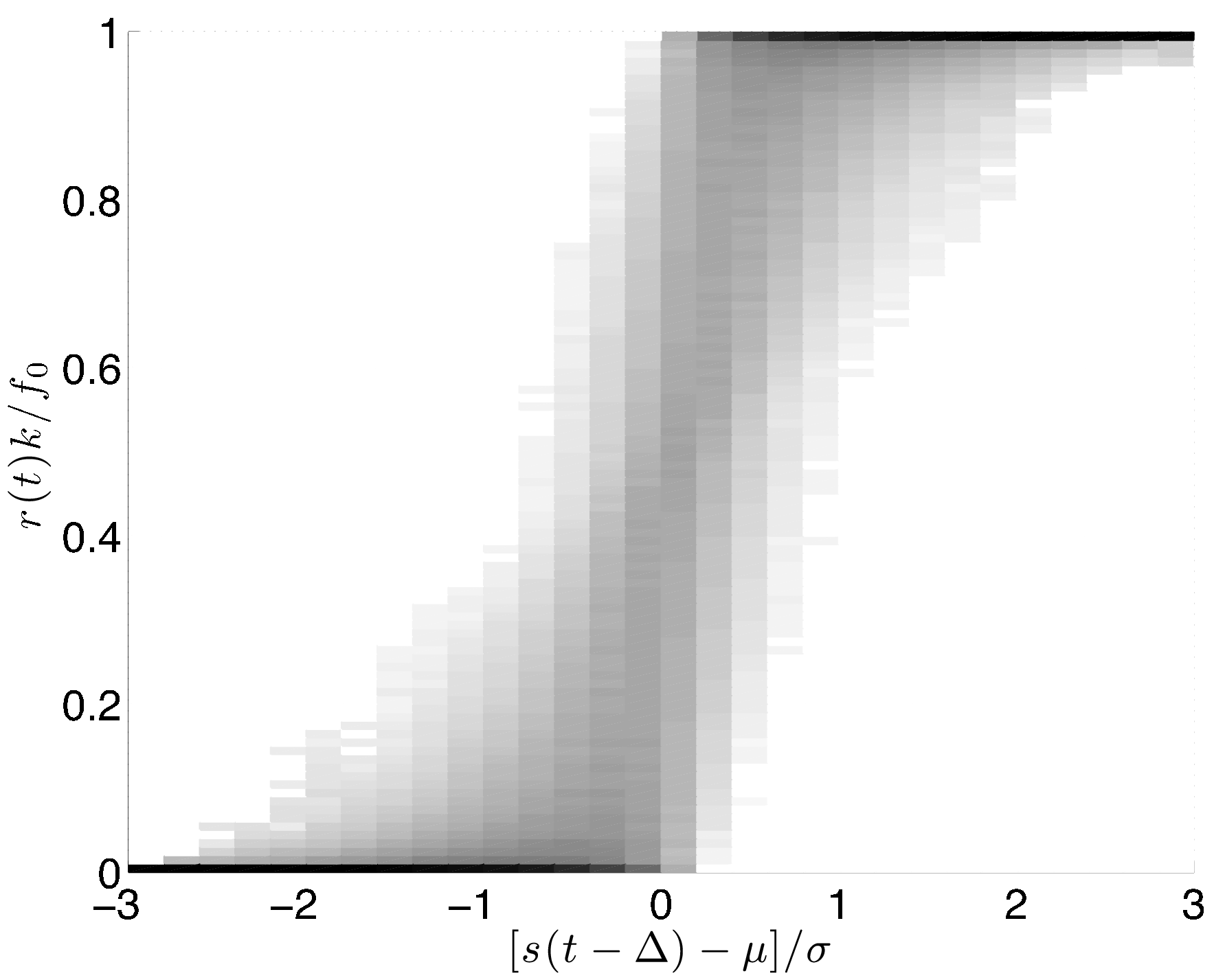}}
\caption{Conditional distribution $P(r(t)|s(t-\Delta))$ for
  $k\tau=10$. The signals are discretized into 30 values in the range
  of $[-3\sigma,+3\sigma]$. For each $s(t-\Delta)$, a histogram of
  $r(t)$ is built with 100 distinct $r$ values. The normalized
  histograms are grey-scale coded as columns on the plot, with dark
  representing the higher conditional probability, $P\sim 1$. We use a
  nonlinear color scale to enhance the plot.  }
\label{conditional}
\end{figure}

In Fig.~\ref{conditional}, we plot the input-output relation for
$k\tau=10$. To emphasize the {\em independence} of the response on
$\sigma$ and hence the gain-compensating nature of the system, we plot
$s$ in units of $\sigma$. A smooth, probabilistic, sigmoidal response
with a width of the transition region $\Delta \sim \sigma$ is clearly
visible. This is because, for a step-function $f$, the value of $r(t)$
depends not on $s(t)$, but on how long the signal has been positive
prior to the current time. In its turn, this duration is correlated
with $s/\sigma$, producing a probabilistic dependence between $r$ and
$s/\sigma$. The latter is manifestly invariant to variance changes.

These arguments make it clear that the fidelity of the response curve
should depend on the ratio of characteristic times of the signal and
the response, $k\tau$. Indeed, as seen in Fig.~\ref{examples}, for
$k\tau\to0$, the response integrates the signal over long times. It is
little affected by the current value of the signal and does not span
the full available dynamic range. At the other extreme of a very fast
response, $k\tau\to\infty$, the system is almost
quasi-stationary. Then the step-nature of $f$ is evident, and the
response quickly swings between two limiting values ($f_0/k$ and 0).

\begin{figure}[b]
\centerline{\includegraphics[width = 6cm]{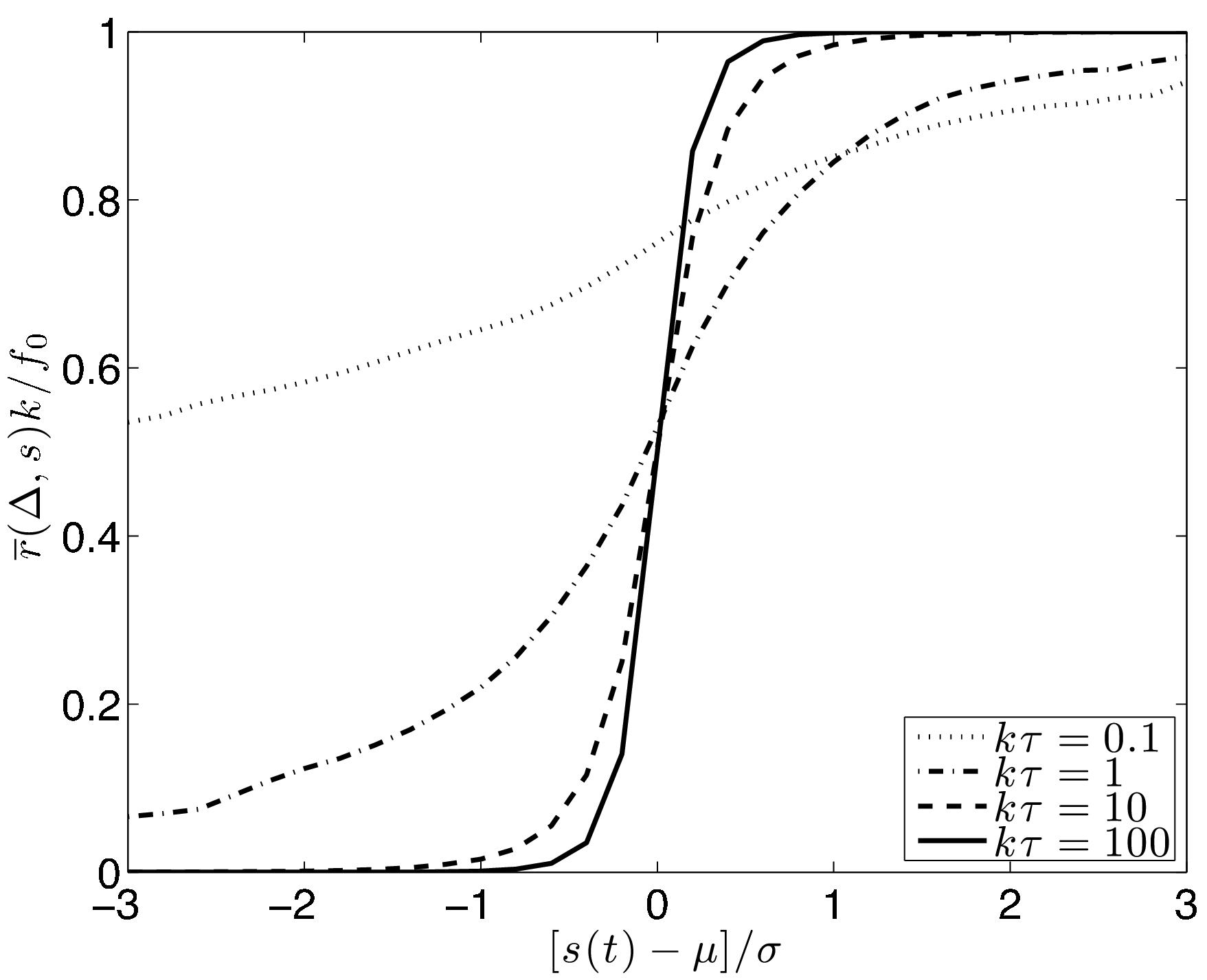}}
\caption{Mean conditional response $\bar{r}(\Delta,s)$ for different
  combinations of the signal and the response characteristic times,
  $k\tau$. }
\label{means}
\end{figure}
\begin{figure}[b]
\centerline{\includegraphics[width = 6cm]{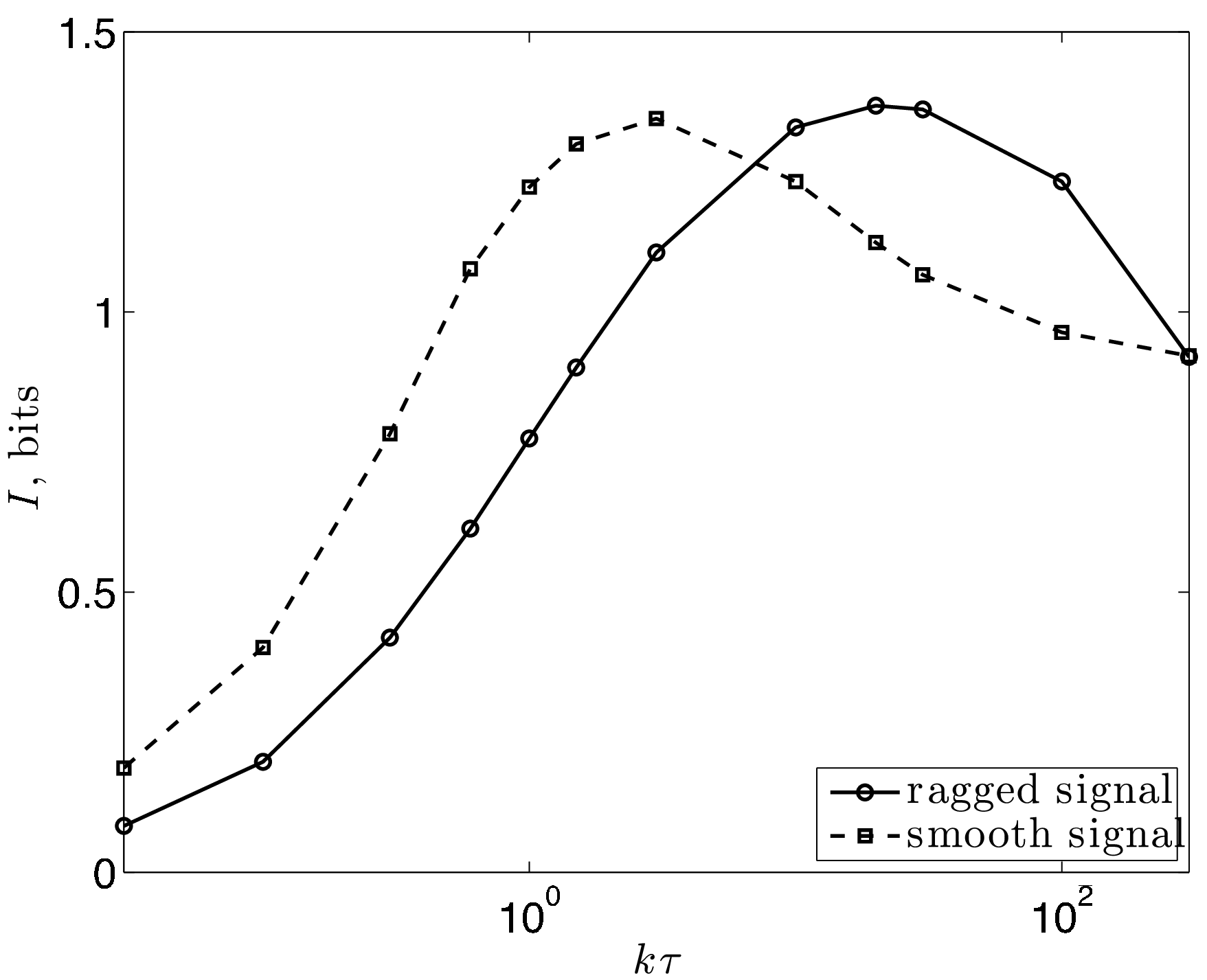}}
\caption{The signal-response mutual information at the optimal
  temporal delay as a function on $k\tau$. The solid line represents
  $I_k[r(t+\Delta),s(t)]$, the information for the Ornstein-Uhlenbeck
  signal, and the maximum of the information here is $I_{\max}=1.37$
  bits, achieved at $k^*\approx20/\tau$.  The dashed line stands for
  the same information for the smoothed signal. It is maximized at
  $k^*\approx 3/\tau$ with $I_{\max}=1.35$ bits. }
\label{mi1}
\end{figure}

We illustrate the dependence of the response conditional distribution
on the integration time in Fig.~\ref{means} by plotting
$\bar{r}(\Delta,s)=\int dr\, r(t+\Delta) P(r(t+\Delta)|s(t))$, the
conditional-averaged response for different values of $k\tau$.
Neither $k\tau\to 0$ nor $k\tau\to\infty$ are optimal for signal
transmission. One expects existence of an optimal $k^*$, for which
most of the dynamic range of $r$ gets used, but the response is not
completely binary. To find this optimum, we evaluate the mutual
information \cite{shannon-49} between the signal and the response at
the optimal delay, $I_k[r(t+\Delta),s(t)]$, as a function of $k\tau$,
cf.~Fig.~\ref{mi1}. A broad maximum in information transmission is
observed near $k^*\approx 20/\tau$, which is not too far from the
quasi-stationary limit. However, $I_{\max}\equiv I_{k^*}=1.37$ bits is
substantially larger than 1. Thus temporal correlations in the
stimulus allow to transmit 37\% more information about it than the
step response would suggest naively. This information is transmitted
in a gain-controlled manner, so that changes in $\sigma$ have no
effect. Similar conclusion should hold for non-step-like $f$, as long
as $f$ is sigmoidal and $\Delta s/\sigma\ll 1$.

{\em Effects of the signal structure.} The observed gain insensitivity
depends only weakly on details of the temporal structure of the
signal. As long as there are autocorrelations, one can use them to
transmit more than one bit about the signal in a gain-independent
fashion using the strong nonlinearity of $f$. To verify this, we
replace the Ornstein-Uhlenbeck signal, Eq.~(\ref{signal}), with its
low-pass filtered version, $s'(t)=1/k\int^tdt's(t')e^{-k(t-t')}$, and
$k$ is the same as in Eq.~(\ref{filter}). This new signal is smoother
and has less structure at high frequencies. We repeat the same
analysis as above to find $\Delta$, estimate the conditional response
distribution, and then evaluate $I_k$, the stimulus-response
information. We find that the maximum information in this case is
$I_{\max}=1.35$ bits, statistically indistinguishable from the
Ornstein-Uhlenbeck case. However, the maximum is now at
$k\tau\approx3$. This is because the smooth signal has a lot fewer
short-lived zero-crossings, and smaller integration times are needed
to approach the extreme values of the response.

{\em Knowing $\sigma$ in a gain-insensitive response.} When
gain-insensitive, the system looses information about the actual
signal variance. This rarely happens in biology. For example, while we
see well at different ambient light levels, we nonetheless know how
bright it is outside. For the fly visual system, it was shown that
variance independence of the response breaks on long time scales. The
signal variance can be inferred from long-term features of the neural
code \cite{brenner-00,fairhall}. Correspondingly, we ask if long term
observation of the response of an approximately gain-controlled
molecular signaling circuit allows to infer the signal variance
$\sigma$.

To this extent, consider $f$ as a narrow sigmoid, with the width of
the crossover region $\Delta s/\sigma\ll 1$. The effect of the
variance on the response is still negligible. For concreteness, we
take $f=f_0[ \tanh((s-\mu)/\Delta s)+1]$.  Consider now the fraction
of time the derivative of the response is near $\max(f)$. This
requires that $r\approx 0$ (so that the degradation, $kr$, is
negligible), but $s$ is already large, $(s-\mu)/\Delta s\gg1$. The
probability of this happening depends on the signal variance and hence
on the speed with which the signal crosses over the threshold
region. Thus one can estimate $\sigma$ by observing a molecular
circuit for a long time and counting how often the rate of change of
the response is large.  While the probability of a large derivative
will depend on the exact shape of $f$, for a signal defined by
Eq.~(\ref{signal}), the statistical error of any such counting
estimator will scale as $\propto \sqrt{\tau/T}$. Hence, the system can
be almost insensitive to $\sigma$ on short time scales, but allow its
determination from long observations.

To verify this, we simulate the signal determined by
Eq.~(\ref{signal}) with the $k\tau=20$, which maximizes the
signal-response mutual information. We calculate the mean fraction of
time $\phi$ when the response derivative is above 80\% of its maximum
value. We further calculate the standard deviation of the fraction
$\sigma_\phi$. We repeat this for signals with various $\Delta
s/\sigma$ and for for experiments of different duration, obtaining a
time-dependence of the $Z$-score for disambiguating two signals with
different variances
$Z=(\phi_2-\phi_1)/\sqrt{\sigma_{\phi_1}^2+\sigma_{\phi_2}^2}$, where
the indeces $1,2$ denote the signals being disambiguated. For example,
for distinguishing signals with $\Delta s/\sigma=1/10$ and $1/20$, we
estimate $Z\approx0.8 (T/\tau)^{0.48\pm 0.04}$, consistent with the
square root scaling (the error bars indicate the 95\% confidence
interval). That is, for $T/\tau$ as little as 10, $Z>2$, and the two
signals are distinguishable. Signals with larger variances are harder
to disambiguate. For example, for $\Delta s/\sigma=1/90$ and $1/100$,
$Z\approx9.4\times10^{-3}(T/\tau)^{0.56\pm0.08}$, and $Z$ crosses 2
for $T\approx15000\tau$.

This long-term variance determination can be performed molecularly in
many different ways. For example, one can use a feedforward incoherent
loop with $r$ as an input \cite{mangan-03}. The loop acts as a
approximate differentiator for signals that change slowly compared to
its internal relaxation times \cite{sontag-10}. The output species of
the loop can then activate a subsequent species by a Hill-like
dynamics, with the activation threshold close to the maximum of the
possible derivative. If this last species degrades slowly, it will
integrate the fraction of time when $dr/dt$ is above the threshold,
providing the readout of the signal variance.

\section{Discussion}

In this article, we have argued that simple molecular circuitry can
respond to signals in a gain-insensitive way without a need for
adaptation and feedback loops. That is, these circuits can be
sensitive only to the signal value relative to its standard
deviation. To make the mechanism work, the signaling system must obey
the following criteria
\begin{itemize}\itemsep 0mm\topsep0mm
\item a nonlinear-linear (NL) response; that is, a strongly nonlinear,
  sigmoidal synthesis function $f$ integrated (linearly) over time;
\item properly matched time scales of the signal and the response dynamics.
\end{itemize}
In addition, the information about the signal variance can be
recovered, for example, if
\begin{itemize}
\item large excursions of the response derivative can be counted over
  long times.
\end{itemize}

Naively transmitted information of only one bit (on or off) would be
possible with a step-function synthesis $f$.  However, the response in
this system is a time-average of a nonlinear function of the
signal. This allows to use temporal correlations in the signal to
transmit more than 1 bit of information for broad classes of
signals. While 1.35 bits may not seem like much more than 1, the
question of whether biological systems can achieve more than 1 bit at
all is still a topic of active research
\cite{ziv-07,tkacik-08}. Similar use of temporal correlations has been
reported to increase information transmission in other circuits, such
as clocks \cite{mugler-10}. In practice, in our case, there is a
tradeoff between variance-independence and high information
transmission through the circuit: a wider synthesis function would
produce higher maximal information for properly tuned signals, but the
information would drop down to zero if $\Delta s\gg\sigma$. It would
be interesting to explore the optimal operational point for this
tradeoff under various optimization hypotheses.

While our analysis is applicable to any molecular system that
satisfies the three conditions listed above, there are specific
examples where we believe it may be especially relevant.  The {\em E.\
  coli} chemotaxis flagellar motor has a very sharp response curve
(Hill coefficient of about 10) \cite{cluzel-motor}. This system is
possibly the best studied example of biological adaptation to the mean
of the signal. However, the question of whether the system is
insensitive to the signal variance changes has not been addressed. The
ultrasensitivity of the motor suggests that it might be. Similarly, in
eukaryotic signaling, push-pull enzymatic amplifiers, including MAP
kinase mediated signaling pathways, are also known for their
ultrasensitivity \cite{goldbeter-81,huang-96,samoilov-05}. And yet
ability of these circuits to respond to temporally-varying signals in
a variance-independent way has not been explored.

We end this article with a simple observation. While the number of
biological information processing systems is astonishing, the types of
computations they perform are limited. Focusing on the computation
would allow cross-fertilization between seemingly disparate fields of
quantitative biology. The phenomenon studied here, lifted wholesale
from neurobiology literature, is an example. Arguably, computational
neuroscience has had a head start compared to computational molecular
systems biology. The latter can benefit immensely by embracing
well-developed results and concepts from the former.

\ack We thank F Alexander, W Hlavacek, and M Wall for useful
discussions in the earlier stages of the work, participants of {\em
  The Fourth International {\em q-bio} Conference} for the feedback,
and F Family for commenting on the manuscript. We are grateful to R
de Ruyter van Steveninck for providing the data for one of the
figures. This work was supported in part by DOE under Contract No.\
DE-AC52-06NA25396 and by NIH/NCI grant No.\ 7R01CA132629-04.

\end{document}